\documentclass[11pt]{article}
\usepackage{moriond,epsfig} %,graphicx

\newcommand{\wpr}{$W^\prime$~}
\newcommand{\wprm}{W^\prime}

\newcommand{\gsm}{g_{_\mathrm{SM}}}  % only use in math mode

\newcommand{\nullchar}{\hspace*{0ex}}

\newcommand{\beq}{\begin{equation}}
\newcommand{\eeq}{\end{equation}}
\newcommand{\bea}{\begin{eqnarray}}
\newcommand{\eea}{\end{eqnarray}}

\begin{document}
%FERMILAB-Conf-03/xxx-T
\vspace*{4cm}
\title{HOW TO RULE OUT LITTLE HIGGS\\ (AND CONSTRAIN MANY OTHER MODELS)
AT THE LHC}

\author{ZACK SULLIVAN}

\address{Theoretical Physics Department, Fermi National
Accelerator Laboratory, Batavia, IL, 60510-0500}

\maketitle
\abstracts{In this talk I describe how to discover or rule out the
existence of \wpr bosons at the CERN Large Hadron Collider as a
function of arbitrary couplings and \wpr masses.  If \wpr bosons are
not found, I demonstrate the 95\% confidence-level exclusions that can
be reached for several classes of models.  In particular, \wpr bosons
in the entire reasonable parameter space of Little Higgs models can be
discovered or excluded in 1 year at the LHC.}

\section{Introduction}

\indent\indent
Significant attention has been paid to the recent class of models of
electroweak symmetry breaking known as ``Little Higgs'' models.  The
purpose of Little Higgs models is to provide a natural mechanism to
cancel quadratic divergences that appear in the calculation of the
Higgs mass without resorting to supersymmetry (cf.\ Ref.\ 1
for a nice review).  The cancellation of divergences occurs by a
clever alignment of vacuua, and the addition of several new particles
--- several scalars, $Z^{\prime}$ and \wpr bosons, and vector-like top
quarks.  While the detailed mass-spectrum and couplings are very
model-dependent, some features generic to all of the models can be
tested to high precision at the Large Hadron Collider (LHC) at CERN.

The key to probing the Little Higgs spectrum is the search for the
\wpr bosons.  For each SU(2) gauge symmetry that is broken there will be a
new massive charged vector-boson with a typical mass~\cite{Hewett:2002px}
\beq
M_{\wprm} < 6 \;\mathrm{TeV}\left[ \frac{m_H}{200 \;\mathrm{GeV}}\right]^2\,.
\eeq
These \wpr bosons will each introduce a new term to the Lagrangian of
the form
\beq
{\cal L} = \frac{g^\prime}{2\sqrt{2}}
V_{ij}\wprm_\mu\bar q^i\gamma^\mu(1-\gamma_5)q^j\,,
\eeq
where $g^\prime = \gsm F(g_1, g_2,\ldots, g_n)$ is the standard-model
SU(2)$_\mathrm{L}$ coupling times a function of the SU(2)$_{\mathrm{L}
i}$ couplings for each broken SU(2).  (For the Littlest Higgs,
$F=g_1/g_2$.~\cite{Han:2003wu}) In addition, there may be couplings to
the other vector bosons, scalars, and the vector-like top quark that
will contribute to the overall width of the \wpr boson. In general
these have a small numerical effect on the width unless $g^\prime$ is
very small.  In that case, the branching fractions to fermions
decrease, however the branching fraction to the final state examined
below can actually increase with the opening of a new channel (see
Ref.\ 4 for details).

An essential constraint on the individual couplings $g_i$ comes from
their relationship to $\gsm$:
\beq
\frac{1}{g_1^2} + \frac{1}{g_2^2} + \cdots + \frac{1}{g_n^2} =
\frac{1}{\gsm^2} \approx \frac{1}{0.427}
\eeq
must hold, which implies $1.02 \gsm < g_{1,2,\ldots} < \sqrt{4\pi}$.
The upper limit is determined by a requirement of perturbativity.
Hence, for all Little Higgs models there will be at least one \wpr
boson with $0.187 < g^\prime/\gsm < 5.34$, and a preference for
$g^\prime/\gsm \sim 1$ in more complicated scenarios.

It was demonstrated in Ref.\ 5 that the best method to look for \wpr
bosons is to search for a resonant mass peak in the
top-quark/bottom-quark decay channel.  A limit in this search method
applies equally to left or right-handed \wpr bosons, whereas the
handedness of the \wpr boson can be determined from the
spin-correlations of the final state.  The model-independent search
for this peak structure can then be translated directly into bounds on
\emph{any} model with \wpr bosons by using the equations in
Ref.\ 5.  The CDF Collaboration has used this method to set lower mass
bounds on \wpr bosons of 536(566)~GeV assuming SM-like
couplings,~\cite{Acosta:2002nu} where decays to right-handed neutrinos
are (not) allowed.  For pure left-handed \wpr bosons (like those that
appear in Little Higgs models), the current best bound is
786~GeV.~\cite{Affolder:2001gr} In this brief summary, I will
demonstrate obtainable exclusion limits at the LHC for arbitrary
couplings and for several specific models, including Little Higgs.

\section{Searching for \wpr bosons at the LHC}

\begin{figure}[tb]
\includegraphics[width=3in]{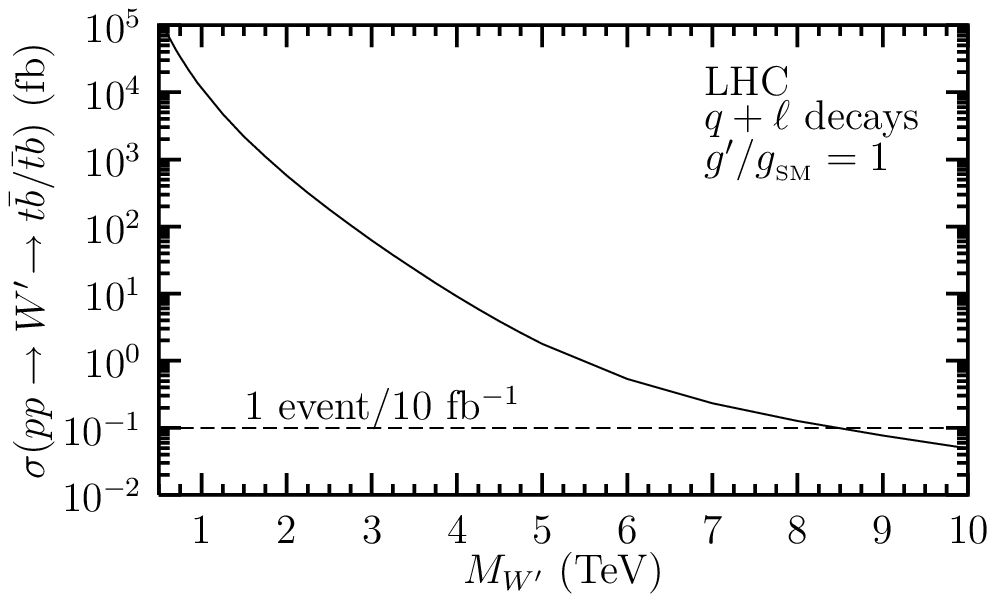}
\includegraphics[width=3in]{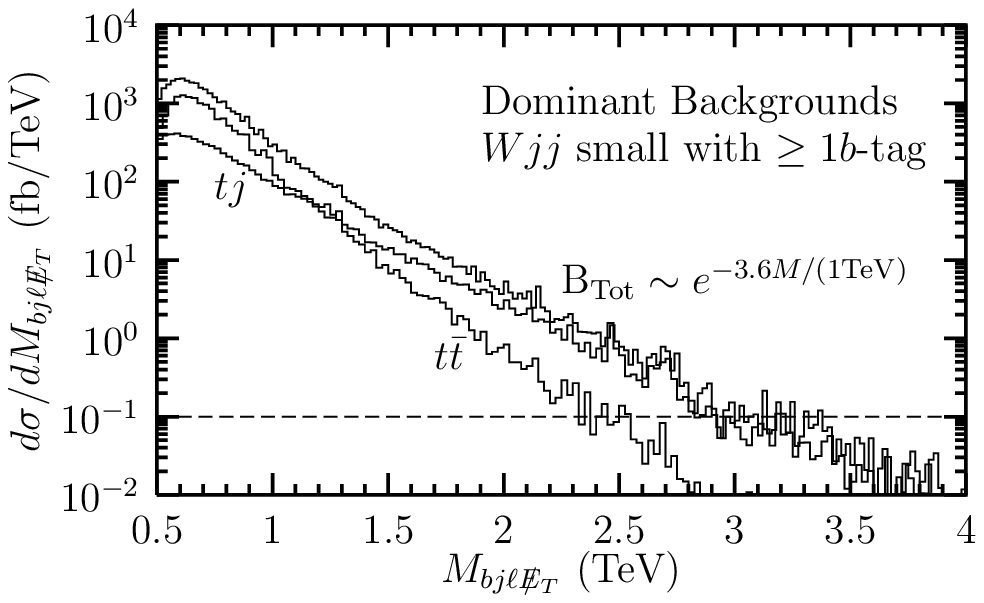}\vspace*{-.4em}\\%
\vspace*{-.6em}
\centerline{\nullchar\footnotesize\hspace*{.25in}(a)\hspace*{3in}(b)}
\caption{(a) Total \wpr cross section at the LHC in (fb).  (b) Total
background and dominant contributions ($t\bar t$ and $tj$) in (fb/TeV)
as a function of the reconstructed $M_{bj\ell\slash\!\!\!\!E_T}$
invariant mass.  The dotted line indicates production of 1 event per
low-luminosity year.\hfill
\label{ZSfig:tot}}
\vspace*{-1em}
\end{figure}

\indent\indent
The $s$-channel production of single top quarks via \wpr bosons can
occur at an extremely large rate at the LHC.  In Fig.\
\ref{ZSfig:tot}a, the cross section for this channel is shown for
SM-like couplings as a function of \wpr mass up to 10~TeV.  The dotted
line denotes the production of 1 \wpr boson decaying into $t\bar b$ or
$\bar tb$ per low-luminosity (10 fb$^{-1}$) year at the LHC.  In
high-luminosity (100 fb$^{-1}$) years there could be 50 \wpr bosons
produced with masses of 10~TeV that decay into this channel.  The
question is, can these be observed over the background?

In order to address this question, a full analysis of the signal and
background has been performed.  The signal is evaluated using
PYTHIA~\cite{Sjostrand:2001yu} run through the SHW detector
simulation~\cite{SHW} with parameters updated to match the ATLAS
detector.~\cite{ATLASTDR} The final state of interest contains a
lepton ($e$ or $\mu$), 2 $b$-jets, and missing energy.  The
backgrounds come from $t\bar t$, $t$-channel single-top-quark
production (i.e.\ $tj$), $Wjj, Wcj, Wb\bar b, Wc\bar c, WZ, Wt$, and
$s$-channel single-top-quark production. As is apparent from Fig.\
\ref{ZSfig:tot}b, the most important of these are $t\bar t$, $tj$, and
$Wjj$.  The cross section for the backgrounds falls exponentially with
$M_{bj\ell\slash\!\!\!\!E_T}$ the reconstructed invariant mass, and
drops to less than one event above 3~TeV.

Unfortunately, the event generators do not currently model the $tj$
background correctly.  Hence, I have used a matrix-element
calculation~\cite{Stelzer:1998ni} normalized to the correct
fully-differential NLO calculation of the $tj$ cross
section.~\cite{Harris:2002md} The resulting jets and leptons are run
through the SHW efficiency routines.  I have checked that the results
are completely insensitive to variations of the detector parameters
within the range of variation quoted in Ref.\ 10.

\begin{table}[tb]
\begin{center}
\caption{Cuts used to reconstruct the $M_{bj\ell\slash\!\!\!\!E_T}$ invariant
mass.  Demand at least 2 jets (with at least 1 $b$-tag), 1 isolated
lepton, and missing energy $\slash\!\!\!\!E_T$.\hfill \label{ZStab:cuts}}
\medskip
\begin{tabular}{ll}
$E_{Tj_1} > \max[ 200~\mathrm{GeV}, \min( 10\%M_{W^\prime},
500~\mathrm{GeV})]$ &
$|\eta_{j_1}| < 2.5$ \\
$E_{Tj_2} > \min( 10\%M_{W^\prime}, 150~\mathrm{GeV})$ &
$|\eta_{j_2}| < 2.5$ \\
$E_{T\ell} > 30~\mathrm{GeV}$ & $|\eta_{\ell}| < 2.5$ \\
$\slash\!\!\!\!E_T > 50~\mathrm{GeV}$&
$100~\mathrm{GeV} < M_{j_2\ell\nu} < 450~\mathrm{GeV}$
\end{tabular}
\end{center}
\vspace*{-1.5em}
\end{table}

In order to extract the signal from the backgrounds I make the cuts
listed in Table~\ref{ZStab:cuts}.  Demanding at least 1 $b$-tag
removes most of the $Wjj$ background.  An additional constraint that
the second-highest-$E_T$ jet combine with the lepton and missing
energy to produce a mass not too far from the top-quark mass can be
useful for eliminating any residual $Wjj$ backgrounds.  In this case,
the neutrino is constructed from the missing energy by using the
$W$-mass constraint, and choosing the smaller of the two possible
rapidities.  Jets are reconstructed using a $k_T$-clustering algorithm
with $R=1$ (similar to a cone size of $0.7$).  For this summary I
chose a mass window which is bounded from below by roughly the \wpr
mass or 3 TeV, whichever is smaller.

\section{Numerical Results}

\begin{figure}[tbh]
\includegraphics[width=3in]{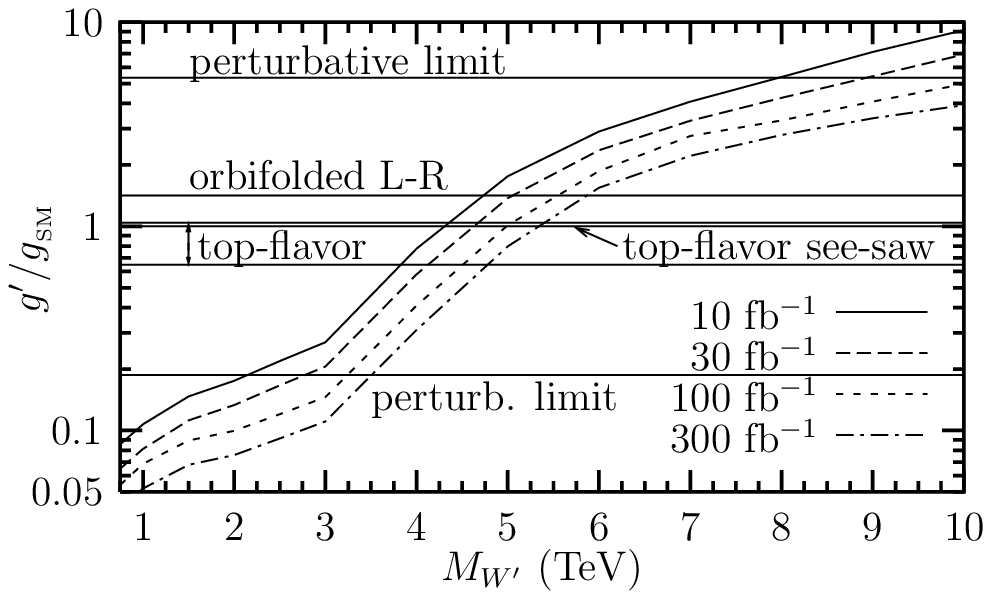}
\includegraphics[width=3in]{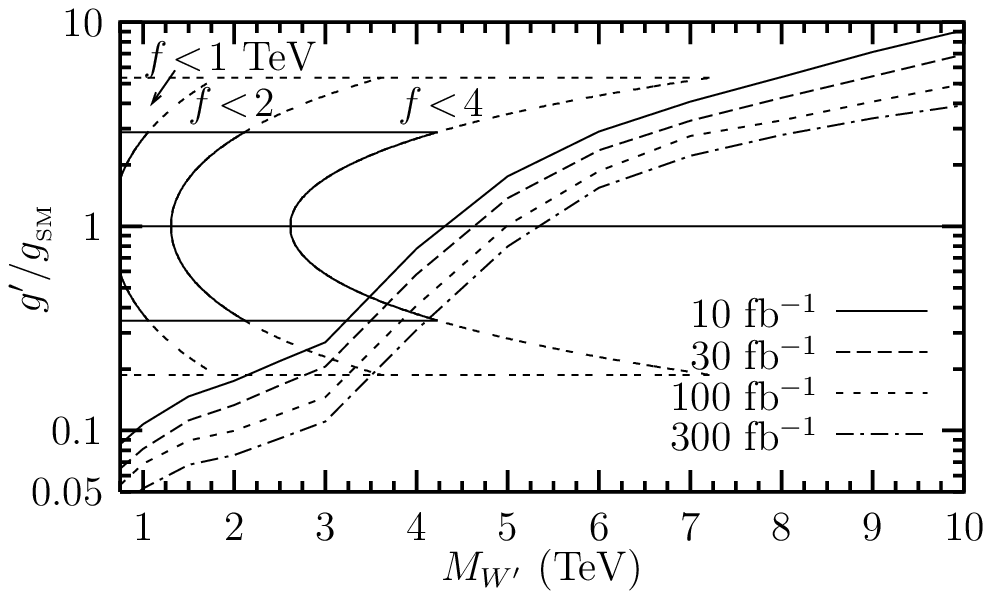}\vspace*{-.4em}\\%
\vspace*{-.6em}
\centerline{\nullchar\footnotesize\hspace*{.25in}(a)\hspace*{3in}(b)}
\caption{95\% confidence-level exclusion reach as a function of \wpr
mass at the LHC for arbitrary $g^\prime/\gsm$.  Superimposed are the
predictions of (a) various classes of perturbative models, and (b)
Little Higgs.  The short dot-dashed contours denote the maximally allowed
parameter space for a given $f$.  The solid contours denote
the perturbative parameter space ($\alpha_i = g_i^2/(4\pi) < 1/\pi\approx
0.32$).\hfill
\label{ZSfig:cmod}}%\label{ZSfig:litlim}
\vspace*{-1em}
\end{figure}

\indent\indent
Nothing in the analysis above is specific to Little Higgs models.  In
fact, these results are generic to all models with a charged
vector-like boson.  For this summary I show results assuming that the
width is composed entirely of decays to fermions (see Ref.\ 4 for the
general case) so that $(g^\prime/\gsm)^2$ is proportional to the
width.~\cite{Sullivan:2002jt} In Fig.~\ref{ZSfig:cmod}a we see the
$95\%$ confidence-level exclusion that can be placed at the LHC as a
function of coupling $g^\prime/\gsm$ and \wpr mass for a few
integrated luminosities.  A \wpr with SM-like couplings can be ruled
out up to $5.5$ TeV!  Overlaid on this plot are the predictions of
several classes of models.  Generic top-flavor
models~\cite{Muller:1996dj,Malkawi:1996fs} predict $0.65 <
g^\prime/\gsm < 1.04$, whereas top-flavor see-saw
models~\cite{He:1999vp} conspire to predict a coupling of
$g^\prime/\gsm = 1$.~\cite{Sullivan:2002jt,Zprod} Another interesting
class of models are orbifolded left-right symmetric
models~\cite{Mimura:2002te} which predict a Kaluza-Klein tower of
right-handed \wpr bosons with an effective coupling $g^\prime/\gsm =
\sqrt{2}$.  In general there is an upper limit on any model with
perturbative couplings of $g^\prime/\gsm \sim 5.34$, and a lower limit
for models with ratios of couplings of $\sim 0.187$.

Now it is time to turn specifically to Little Higgs.  Little Higgs is
one of the theories that is supposed to be perturbative at all stages,
and hence has the absolute limits on the couplings quoted before.
However, there is an additional relationship between $M_\wprm$ and $f$
the pseudo-Goldstone boson decay constant.  To find upper bounds on the
allowed \wpr mass it is sufficient to look at the Littlest Higgs,
where the relationship is
\beq
M_\wprm \approx \frac{f}{2}\sqrt{g_1^2 + g_2^2} \,.
\eeq
Solving for $g^\prime$ in terms of $f$ and the mass leads to the
maximally allowed region of parameter space shown by the triple-dashed
contours of Fig.~\ref{ZSfig:cmod}b.

When examining Fig.~\ref{ZSfig:cmod}b it should be questioned whether
the theory is really perturbative if one of the couplings is
$\sqrt{4\pi}$.  A more reasonable perturbative bound of $\alpha_i =
g_i^2/(4\pi) < 1/\pi \approx 0.32$ is shown via solid contours in
Fig.~\ref{ZSfig:cmod}b.  The figure stops for $f=4$~TeV, since it
becomes increasing unnatural for $f$ to be larger than 1 TeV in Little
Higgs scenarios.~\cite{Hewett:2002px} However, even $f$ as large as
6--8~TeV can be mostly covered in the central perturbative region
($g^\prime/\gsm \sim 1$)which is favored for more complicated models.
The conclusion to be drawn is that the \wpr bosons appearing in Little
Higgs models should either be seen or excluded in the first year of
running at the LHC.\vspace*{-1em}

\section*{Acknowledgments}

\indent\indent
This work is supported by the U.~S.~Department of Energy under
contract No.~DE-AC02-76CH03000.

\end{document}